\documentclass[11pt,twoside]{article}  % Leave intact
\usepackage{apn3conf}

\begin{document}   % Leave intact

%-----------------------------------------------------------------------
%		            Paper Title 
%-----------------------------------------------------------------------
% Enter the title of the paper.
%
% EXAMPLE: \title{A Breakthrough in Planetary Nebulae Research}
%
% If your title is so long as to fill the page header when you print it,
% then please supply a short form as a \titlemark.
%
% EXAMPLE:
%  \title{Magnetic Fields in the Shaping of Planetary Nebulae:
%         Observations Meet Theory}
%  \titlemark{Magnetic Fields in Planetary Nebulae}
%

\title{HD 161796 (IRAS 17436+5003)}
%\titlemark{ }

%-----------------------------------------------------------------------
%		          Authors of Paper
%-----------------------------------------------------------------------
% Enter the authors followed by their affiliations.  The \author and
% \affil commands may appear multiple times as necessary.  List each
% author by giving the first name or initials first followed by the
% last name.  Authors with the same affiliations should grouped
% together. 
%
% Try to limit the front matter to no more than three \author
% commands.  Group authors with the same affiliations.  Too many
% \author commands fills the first page of the paper with little
% actual text.

\author{Toshiya Ueta}
\affil{Royal Observatory of Belgium, Avenue Circulaire 3, B-1180, Belgium}

%-----------------------------------------------------------------------
%			 Contact Information
%-----------------------------------------------------------------------
% This information will not appear in the paper but will be used by
% the editors in case you need to be contacted concerning your
% submission.  Enter your name as the contact along with your email
% address.

\contact{Toshiya Ueta}
\email{ueta@oma.be}

%-----------------------------------------------------------------------
%		      Author Index Specification
%-----------------------------------------------------------------------
% Specify how each author name should appear in the author index.  The 
% \paindex{ } should be used to indicate the primary author, and the
% \aindex for all other co-authors.  You MUST use the following
% syntax: 
%
% SYNTAX:  \aindex{LASTNAME, F. M.}
% 
% where F is the first initial and M is the second initial (if
% used).  This guarantees that authors that appear in multiple papers
% will appear only once in the author index.  
%
% EXAMPLE: \paindex{Crabtree, D.}
%          \aindex{Manset, N.}
%          \aindex{Veillet, C.}
%
% NOTE: this information is also used to build the author list that
% appears in the table of contents.  Authors will be listed in the order
% of the \paindex and \aindex commmands.
%

\paindex{Ueta, T.}

%-----------------------------------------------------------------------
%                     Author list for page header
%-----------------------------------------------------------------------
% Please supply a list of author last names for the page header. in
% one of these formats:
%
% EXAMPLES:
% \authormark{LASTNAME}
% \authormark{LASTNAME1 \& LASTNAME2}
% \authormark{LASTNAME1, LASTNAME2, ... \& LASTNAMEn}
% \authormark{LASTNAME et al.}
%
% Use the "et al." form in the case of seven or more authors, or if
% the preferred form is too long to fit in the header.

\authormark{Ueta}

%-----------------------------------------------------------------------
%			Subject Index keywords
%-----------------------------------------------------------------------
% Enter up to 6 keywords describing your paper.  These will NOT be
% printed as part of your paper; however, they will be used to
% generate an object index and a subject index for the proceedings.  
% There is no standard list,  however, individual object names are
% encouraged and one or two word descriptions of the topics (e.g.MHD, 
% ionized gas) are useful. 
%
% EXAMPLE:  \keywords{NGC 7027, AFGL 2688, HD 161796, binary stars,
%                      dust,  molecular gas}

\keywords{HD 161796, IRAS 17436+5003, mass loss, dust, circumstellar shell}

%-----------------------------------------------------------------------
%			       Abstract
%-----------------------------------------------------------------------
% Type abstract in the space below.  Consult the User Guide and Latex
% Information file for a list of supported macros (e.g. for typesetting 
% special symbols). Do not leave a blank line between \begin{abstract} 
% and the start of your text.

\begin{abstract}          % Leave intact
Past imaging observations of a post-AGB star, HD 161796, have probed 
different parts of its circumstellar shell which correspond to 
different epochs of the star's AGB mass loss history.   
While the overall structure of the shell can be described by an 
axisymmetric model consisting of three layers of characteristic 
structure, the mass distribution in the shell appears to rotate 
its axis of symmetry in a continuous manner.
It is thus interesting to observe the halo region of the
shell using from far-IR to sub-mm wavelengths and increase the 
``time resolution'' during this critical epoch of the star's 
mass loss history.
\end{abstract}

%-----------------------------------------------------------------------
%			      Main Body
%-----------------------------------------------------------------------
% Place the text for the main body of the paper here.  You should use
% the \section command to label the various sections; use of
% \subsection is optional.  Significant words in section titles should
% be capitalized.  Sections and subsections will be numbered
% automatically. 

\section{The Object}

HD 161796 (IRAS 17436+5003) is a low metallicity oxygen-rich post-AGB
star of F3Ib located at high Galactic latitude (e.g., Hrivnak et al.\
1989).  
Due to the star's relative proximity ($\sim 1$ kpc), its circumstellar
shell has been relatively well studied at various wavelengths in order
to understand its mass loss history.

\section{The Outer Shell}

The circumstellar shell around HD 161796 was marginally resolved by 
{\sl IRAS} at 100 $\mu$m, and the {\sl IRAS} map seems to show a
spherically symmetric shell. 
Using {\sl ISO} linear scan observations across the shell at 90 and 160
$\mu$m, Speck et al.\ (2001) have recently shown that the shell is 
indeed extended out to about 400$^{\prime\prime}$.
Even though the {\sl ISO} scan was done only in one direction, it seems
reasonable that this cold ($\sim 20$ K) dust shell possesses generally
spherically symmetric structure. 
This parsec-sized, nearly spherical cold dust shell corresponds to the
part of the circumstellar shell that was created by an ancient AGB mass
loss roughly $2 \times 10^{5}$ years ago (Fig.\ \ref{fig1}, left).  

\section{The Inner Shell}

\begin{figure}
\epsscale{1.0}
\plotone{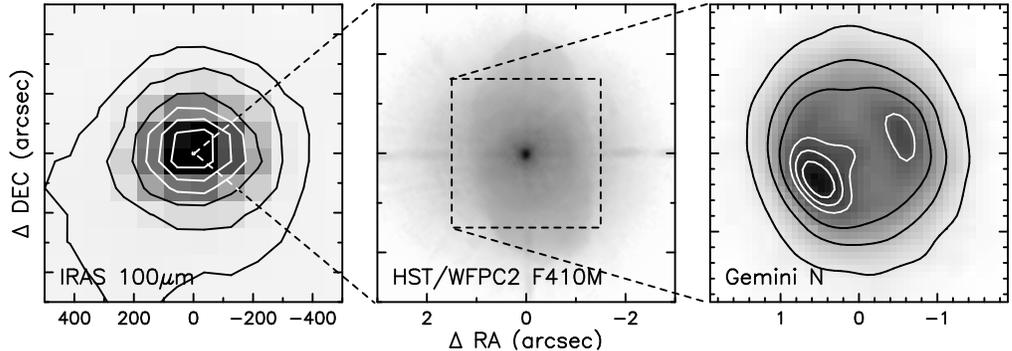}
\caption{%
The shell structure of HD161796.
[Left] The outer spherical shell of cold dust seen in {\sl IRAS\/} 100
 $\mu$m. 
[Middle] The spheroidal mid-shell seen in optical reflection by {\sl
 HST\/} (Ueta et al.\ 2000).
[Right] The inner dust torus seen in Gemini N-band (Gledhill et al.\
 2003). 
Dotted line indicates the corresponding FOV.} \label{fig1}
\end{figure}

While the outer shell represent the old AGB mass loss history, the inner
shell is a manifestation of the most recent mass loss history that may
provide clues for the latest structure formation in the circumstellar
shell.  
By directly probing the distribution of warm ($\sim 100 - 200$ K) dust
grains, Skinner et al.\ (1994) have revealed a two-peaked structure in
the emission core in their deconvolved 12 $\mu$m image.
Most recently, Gledhill et al.\ (2003) have taken N and Q band images
of the object using Gemini-N and resolved the two-peaked structure in 
the mid-IR emission core (Fig.\ \ref{fig1}, right). 
The two-peaked mid-IR structure is typically interpreted as
limb-brightened edges of a dust torus (or an equatorially enhanced dust
distribution) embedded in the innermost region of the circumstellar shell.

The dust shell structure can also be investigated indirectly through
dust-scattered star light in the optical and near-IR. 
As part of their {\sl HST} imaging survey of PPNs, Ueta et al.\ (2000)
have detected an elliptically elongated shell surrounding the mid-IR
core (Fig.\ \ref{fig1}, middle).
These non-spherically symmetric structure of the shell seem to have
developed during the last $\sim 500$ years before the mass loss was
terminated nearly 300 years ago (Meixner et al.\ 2002).

\section{The Overall Shell Structure}

\begin{figure}
\epsscale{1.0}
\plottwo{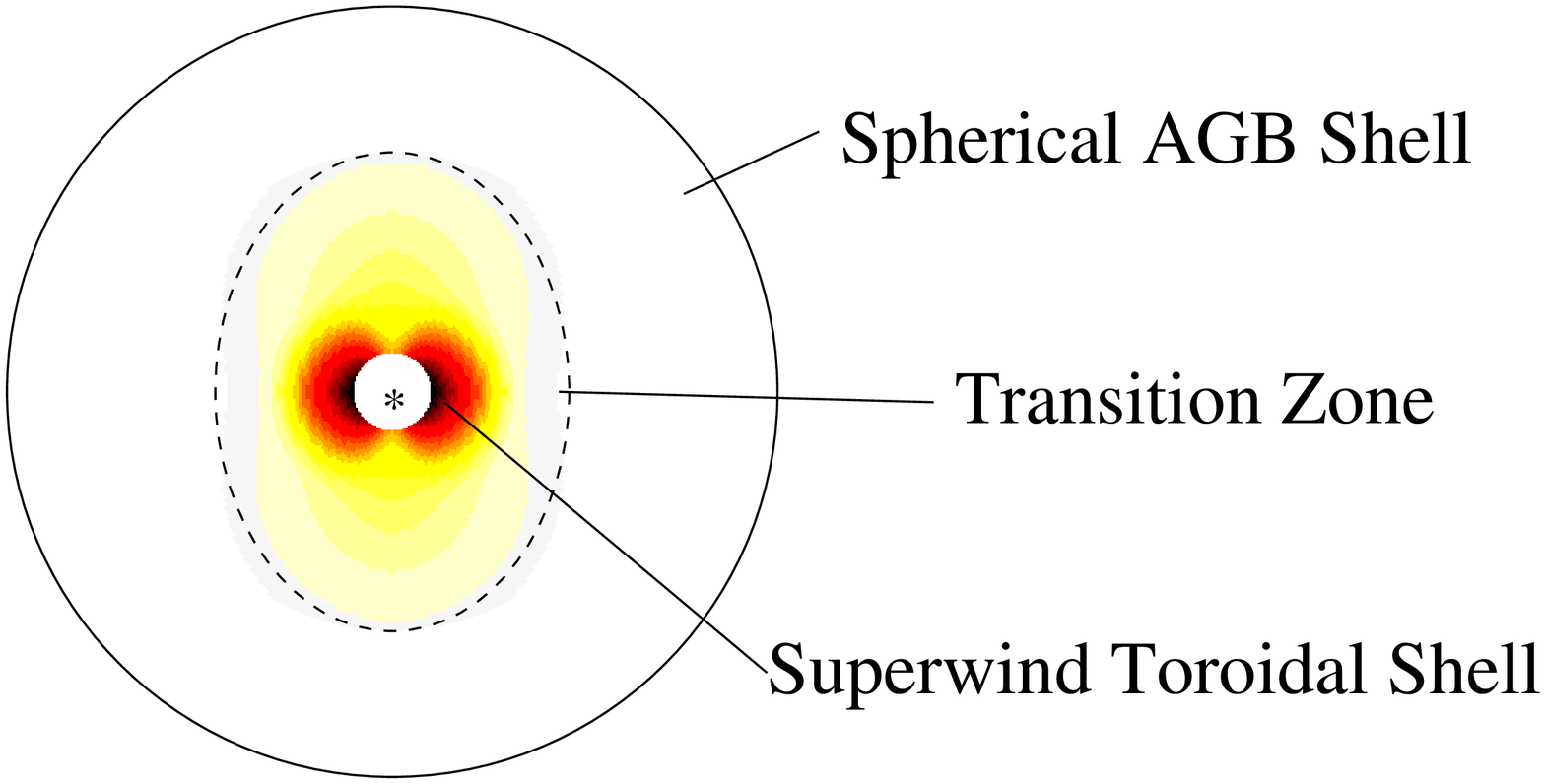}{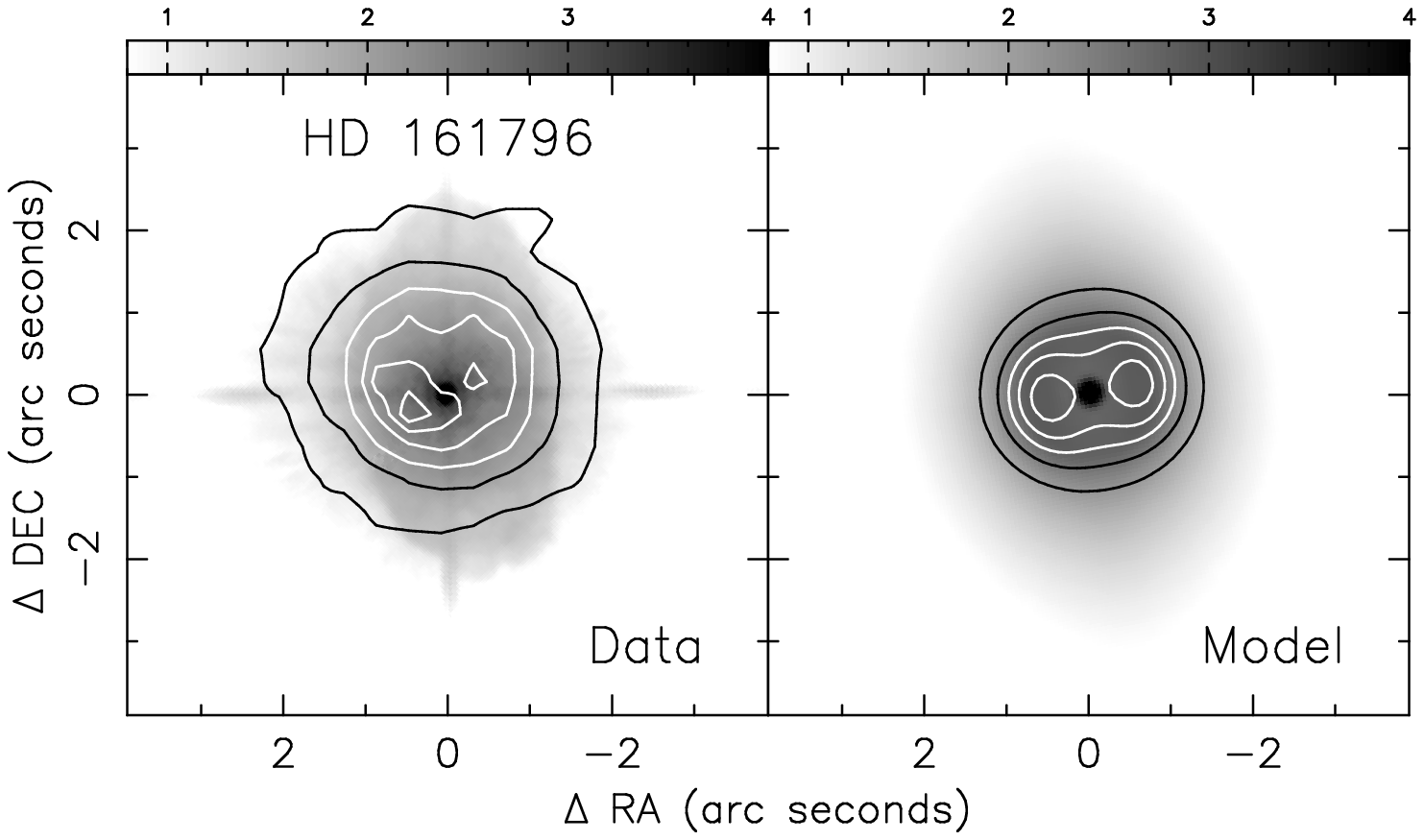}
\caption{%
The ``layered shell'' model for HD 161796.
[Left] A schematic representation of the model
(Ueta et al.\ 2003).
[Right] 2.5-D radiative transfer model of HD 161796 showing the reflection
 nebula in gray scale and the mid-IR nebula in contours
(Meixner et al.\ 2002).} \label{fig2}
\end{figure}

These images of the shell at various wavelengths probe different parts
of the shell, and hence, can be thought of as ``time slices'' that would
help to reconstruct the entire AGB mass loss history of the star.
Based on the shell structure we see at different epochs, the overall
shell structure seems to be explained by a ``layered shell'' model, in
which the shell consists of three layers that represent 
different epochs of mass loss (Fig.\ \ref{fig2}, left). 

The outermost layer of cold dust corresponds to the part of the shell
created by the earliest AGB mass loss, in which the wind was spherically
symmetric. 
The innermost layer of warm dust, on the other hand, corresponds to the 
part of the shell generated by equatorially enhanced (or
axisymmetric/toroidal) mass loss at the end of the AGB phase.
In addition, the mid-layer seen in the dust-scattered star light
corresponds to one particular epoch during which the shell geometry  
transforms from spherical to axisymmetric (toroidal) symmetry.

2.5-dimensional dust radiative transfer calculations have been performed
to test this layered shell model (Meixner et al.\ 2002; Ueta et al.\ 2003).
The best-fit model has successfully reproduced the spectral energy
distribution from UV to far-IR and characteristic morphologies at
various wavelengths from the optical through mid-IR (Fig.\ \ref{fig2},
right).

%, while attributing
%the morphological dichotomy (elliptical vs.\ bipolar) in the post-AGB
%shells to the optical depth of the shell (plus an auxiliary effects of
%inclination). 

\section{The ``Twist'' in the Shell}

Although the overall shell structure can be understood by the layered
shell model consisting of three distinct structures, the model is not
enough to explain the detailed structure of the shell, especially the
apparent deviation from axisymmetry.
If we define the axis of the shell based on the toroidal structure in
the mid-IR core, the axis would have the position angle of roughly 
30$^{\circ}$. 
However, the elliptical structure of the dust-scattered reflection
nebulosity would delineate the axis of symmetry oriented at roughly
0$^{\circ}$ position angle. 
By comparing these two ``time slices'' (i.e., mid-IR and optical images 
representing different epochs of mass loss), this apparent ``shift'' of
the shell orientation seems to occur rather abruptly.

\begin{figure}
\epsscale{0.7}
\plotone{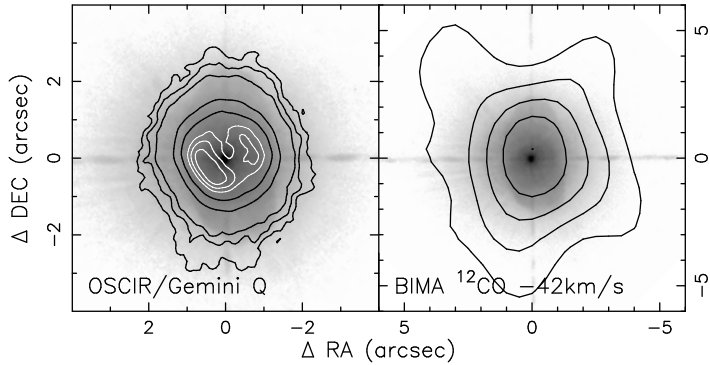}
\caption{%
``Twist'' of the shell axis in HD 161796.
[Left] The mid-IR torus (white contours) does not exactly align with
 the optical reflection nebulosity (gray scale), which  is
 aligned with the mid-IR halo (black contours).
[Right] CO gas distribution seems to indicate continuous ``twisting'' of
 the shell structure at far radii (contours).} \label{fig3}
\end{figure}

However, recent mid-IR observations have captured dust emission of the
innermost core and of the low emission halo regions in a single image. 
So, the axis of symmetry now appears to shift continuously in the
counter-clockwise direction over the course of the star's mass loss
history (Gledhill et al.\ 2003; Fig.\ \ref{fig3}, left). 
Interestingly, if one follows the density distribution further away from
the central star probing even colder parts of the shell through CO gas
in the mm-wavelengths, the twisting of the shell seems to be continuing
for quite some time (roughly 1000 years; Fong et al.\ 2003;
Fig.\ \ref{fig3}, right).

\section{Summary}

As we have seen, imaging of the circumstellar shells at various
wavelengths is equivalent to taking ``snapshots'' of the mass
distribution at various epochs of the mass loss history.
This method has been quite effective to observationally establish the
geometrical transition of the circumstellar shells.
Recent imaging of HD 161796 at the mid-IR has probed the outer shell  
of cold dust and has suggested that the axis of symmetry rotates
continuously while the shell morphology assumes more
equatorially enhanced structure.

Therefore, it is very interesting to increase the ``time resolution'' of
the early to intermediate phases of the mass loss history represented by
rather cold (around $20 - 100$ K) part of the dust shells (corresponding
to the mass loss epochs on the order of $10^{3}$ years ago). 
With far-IR to sub-mm telescopes/missions such as {\sl SIRTF}, 
{\sl SOFIA}, {\sl ASTRO-F}, {\sl VLTI}, {\sl Herschel}, and {\sl SMA},
we will have a plenty of opportunity to observe these critical regions
of the circumstellar shells at higher resolution and with higher
sensitivities.
We can then follow the early phases of the mass loss history
and address how spherically symmetric circumstellar shells 
develop the equatorially-enhanced structure.
This line of research may also provide clues to explain how elliptical
and bipolar PPNs assume their respective structure by comparing
their early mass loss history.

% You can also add an acknowledgments section as indicated below.

%\acknowledgments

%-----------------------------------------------------------------------
%			      References
%-----------------------------------------------------------------------
% List your references below within the reference environment
% (i.e. between the \begin{references} and \end{references} tags).
% Each new reference should begin with a \reference command which sets
% up the proper indentation.  Observe the following order when listing
% bibliographical information for each reference:  author name(s),
% publication year, journal name, volume, and page number for
% articles.  Note that many journal names are available as macros; see
% the User Guide for a listing "macro-ized" journals.   
%
% Note the following are some of the tricks that can be used:
%
%   o  \& is used to format an ampersand symbol (&).
%   o  \'e puts an accent grave over the letter e.  See the User Guide
%      for details on formatting special characters.  
%   o  "\ " after a period prevents LaTeX from interpreting the period 
%      as an end of a sentence.
%   o  \aj is a macro that expands to "Astron. J."  See the User Guide
%      for a full list of journal macros
%

% Do not place any material after the references section

\end{document}